\documentclass[journal]{IEEEtran}
\usepackage{cite}
\usepackage{hyperref}

\ifCLASSINFOpdf
  \usepackage[pdftex]{graphicx}
  \DeclareGraphicsExtensions{.pdf,.jpeg,.png}
\else
\fi
\usepackage{amsmath}
\usepackage{amsfonts}
\usepackage{multirow}
\usepackage{makecell} 

  \usepackage[caption=false,font=footnotesize]{subfig}
 \usepackage{dblfloatfix}

\usepackage{academicons}
\usepackage{xcolor}
\newcommand{\orcid}[1]{\href{https://orcid.org/#1}{\textcolor[HTML]{A6CE39}{\aiOrcid}}}

\begin{document}
%
\title{Unsupervised foreign object detection based on dual-energy absorptiometry in the food industry}
%
%
%

\author{Vladyslav~Andriiashen, 
        Robert~van~Liere, 
        Tristan~van~Leeuwen, 
        Kees~Joost~Batenburg
\thanks{V.~Andriiashen, R.~van~Liere, T.~van~Leeuwen, and K.~J.~Batenburg are with Centrum Wiskunde \& Informatica, Science Park 123, 1098 XG Amsterdam, The Netherlands (email: vladyslav.andriiashen@cwi.nl, robert.van.liere@cwi.nl, t.van.leeuwen@cwi.nl, joost.batenburg@cwi.nl)}
\thanks{R.~van~Liere is with Faculteit Wiskunde en Informatica, Technical University Eindhoven, Groene Loper 5, 5612 AZ Eindhoven, The Netherlands}
\thanks{T.~van~Leeuwen is with Mathematical Institute, Utrecht University, Budapestlaan 6, 3584 CD Utrecht, The Netherlands}
\thanks{K.~J.~Batenburg is with Leiden Institute of Advanced Computer Science, Niels Bohrweg 1, 2333 CA Leiden, The Netherlands}
}

%
%

\markboth{IEEE TRANSACTIONS ON IMAGE PROCESSING}%
{}
%



\maketitle

\begin{abstract}
X-ray imaging is a widely used technique for non-destructive inspection of agricultural food products. One application of X-ray imaging is the autonomous, in-line detection of foreign objects in food samples. Examples of such inclusions are bone fragments in meat products, plastic and metal debris in fish, fruit infestations.

This article presents a processing methodology for unsupervised foreign object detection based on dual-energy X-ray absorptiometry (DEXA). A foreign object is defined as a fragment of material with different X-ray attenuation properties than those belonging to the food product. A novel thickness correction model is introduced as a pre-processing technique for DEXA data. The aim of the model is to homogenize regions in the image that belong to the food product and enhance contrast where the foreign object is present. In this way, the segmentation of the foreign object is more robust to noise and lack of contrast.

The proposed methodology was applied to a dataset of 488 samples of meat products. The samples were acquired from a conveyor belt in a food processing factory.  Approximately 60\% of the samples contain foreign objects of different types and sizes, while the rest of the samples are void of foreign objects. The results show that samples without foreign objects are correctly identified in 97\% of cases, the overall accuracy of foreign object detection reaches 95\%.
\end{abstract}


%
\IEEEpeerreviewmaketitle

\section{Introduction}
%
%
%
%

\IEEEPARstart{A}{}gricultural food products naturally vary in their detailed internal structure. Individual samples can be analyzed to assess product quality, predict maturity state and minimize waste. To facilitate early detection of health risks, it is crucial to apply an inspection procedure capable of detecting foreign object inclusions\cite{chen2001multiresolution, kwon2008real, kotwaliwale2014x} and contaminations\cite{van2016segmentation, chuang2011automatic}. This task can be performed for an individual sample by a human expert. However, a manual inspection can not provide reasonable speed in high-throughput cases, such as product processing in a factory.

X-ray imaging is a widely used technique for non-dectructive, in-line inspection of agricultural products\cite{el2019applications}. It is commonly applied to food samples while they are being processed on a conveyor belt on the factory floor. One of the important applications of X-ray is the automatic detection of foreign object inclusions that might appear in food products. Examples of such objects are bone fragments in a chicken filet, plastic debris and bones in fish, infestation in fruits. One of the well-known approaches in foreign object detection is to acquire two dual-energy X-ray absorptiometry (DEXA)\cite{lopez2018rapid} projections with different values of the X-ray tube voltage. This method is commonly used in medical X-ray imaging for contrast agent detection and body composition analysis. 

In-line foreign object detection in food samples on a conveyor belt possesses three major challenges for DEXA analysis. Firstly, high-throughput X-ray acquisition leads to a significant noise level that will greatly impact the detection process. Noise reduction methods have to be applied to reduce this effect. Secondly, typical foreign objects in the food industry have similar X-ray attenuation properties as food samples, resulting in low contrast between foreign object and main object. Contrast enhancing methods have to be applied to mitigate this effect. Thirdly, a variance of thickness of the main object causes ambiguities when detecting a foreign inclusion in a sample. For example, a thin sample of a chicken filet with a bone fragment might have a very similar dual-energy measurements as a thicker sample of a filet object without a bone. Hence, a thickness correction method should be taken into account
to mitigate this effect. 

The main contribution of this article is a novel approach to DEXA image pre-processing. For low contrast foreign objects, it is crucial to analyze how the ratio between two projections acquired with different voltages depends on the thickness of the object. This effect is caused by the polychromatic spectrum of the X-ray tube and the unknown shape of the sample. The correlation between two intensities for different voltages is not the same for the defect and the product sample, and this difference can be utilized to distinguish between them. The goal of the DEXA pre-processing is to create an image where the main object has an average intensity of zero whereas the foreign object deviates significantly from zero.

This article presents a three step data processing methodology that uses the DEXA pre-processing model in combination with an active contour algorithm and parameterizable foreign object detection criteria. The methodology is optimized to achieve high detection rates on samples with foreign object inclusions and, perhaps more importantly for industrial applications, achieve low detection rates on samples without inclusions. Although the results are targeted towards bone fragments in chicken filets, the processing methodology is generic and can be used to analyse the performance of various industrial scenarios of foreign object detection on a conveyor belt. The DEXA pre-processing is based on general concepts of X-ray measurements and can be applied to other materials. The active contour model is chosen for the segmentation in this work. However, the outcome of the pre-processing step can be used as an input for other segmentation algorithms, including machine learning based methods.


\section{Related work}

Non-destructive study of the products is an important topic for the food industry. It is applied to a variety of food samples \cite{du2019x}, and every type of object has different details and typical defects to detect. X-ray imaging can be used for the detection of grain infection, fruit infestation, bone detection in the fish and poultry industry. X-ray CT is of great interest since it enables volume reconstruction and detection of defects based on the internal structure of the product. However, this approach requires a significant time for data acquisition and reconstruction. Discrete tomography based on limited-angle data \cite{pereira2017inline} can be introduced to balance acquisition time and reconstruction quality. This paper will concentrate on the radiography approach since it can provide the fastest inspection.

Foreign object detection based on a single projection has been studied for different types of food, such as poultry and apples. Multiple algorithms for fruit inspection rely on the shape knowledge that can be estimated with a certain accuracy \cite{van2019combination}. However, knowledge of the shape of the product is not necessary if the foreign object has significantly different X-ray absorption properties. Several algorithms can be applied to enhance foreign object detection by utilizing conventional filtering\cite{mery2011automated}, local contrast enhancement \cite{chen2001multiresolution} and local adaptive thresholding \cite{mathanker2010local}. This approach relies on the assumption that an absorption gradient on the border of the defect can be distinguished from gradients in the main object.

The addition of the second projection with a different tube voltage for better defect detection is a concept that is widely used in medicine. It is applied to a body fat measurement \cite{vasan2018comparison} that determined percentages of different tissues in the human body based on their absorption rates corresponding to different tube voltages. This problem may be similar to some types of food inspection (assessment of fat level), but defect detection focuses on small inclusions of different objects. Furthermore, in many applications, material identification is performed using dual-energy CT\cite{martin2015learning}. This approach is more accurate than DEXA because attenuation properties are analyzed for a small region of the internal structure of the object. If the data are limited to a single projection, the measured attenuation distribution depends on both material properties and the unknown shape of the object (thickness along the ray trajectory). In \cite{international2011iaea}, this effect was explained by beam hardening and corrected with a system calibration. 

In the poultry industry, an addition of a laser was considered to obtain a thickness profile of the studied object \cite{tao2000thickness, gleason2002automatic}. Knowledge of the exact thickness profile helps to predict an absorption distribution for the main object if it is homogeneous. Thus, the presence of the defect can be detected by simple thresholding after subtracting the expected distribution from the measured absorption signal. In this study, only X-ray equipment will be used to perform detection, and no additional sources of information will be used.

Active contours methods can be used for foreign object detection if there is a visible boundary separating a defect and the main object. The level-set methods of Osher and Sethian were used for a fan bone location in the chicken samples \cite{vachtsevanos2000fusion} based on the combination of X-ray and color images. The main downside of many active contour models is that they rely on the edges to perform segmentation. With a high noise level, any edge information becomes unreliable since noise deviations are bigger than a natural absorption gradient that would be observed on a noiseless image. The Chan-Vese energy equation \cite{chan2001active} was used to partition an image into two phases without relying on edge detection. This algorithm is unsupervised and does not require any prior knowledge or machine learning to work.

In this work, the inspection procedure is evaluated based on the detection rate and not segmentation accuracy. Typical studies of the algorithms concentrate on the images with a defect and estimate the accuracy of segmentation. However, the detection rate is more important for most industry applications since the majority of the samples is expected to be without a defect. Such study was performed, for example, in \cite{van2020nondestructive} for pear fruit inspection. Methods to distinguish between bones and no-bones for different patches of fish images were proposed in \cite{mery2011automated}. An algorithm with a good detection rate might show suboptimal segmentation accuracy for samples with a foreign object. However, a good inspection procedure requires a balance in performance on normal and defected samples.

\section{Methods}

\subsection{General methodology}

\begin{figure*}[!t]
\centering
\includegraphics[width=0.8\linewidth]{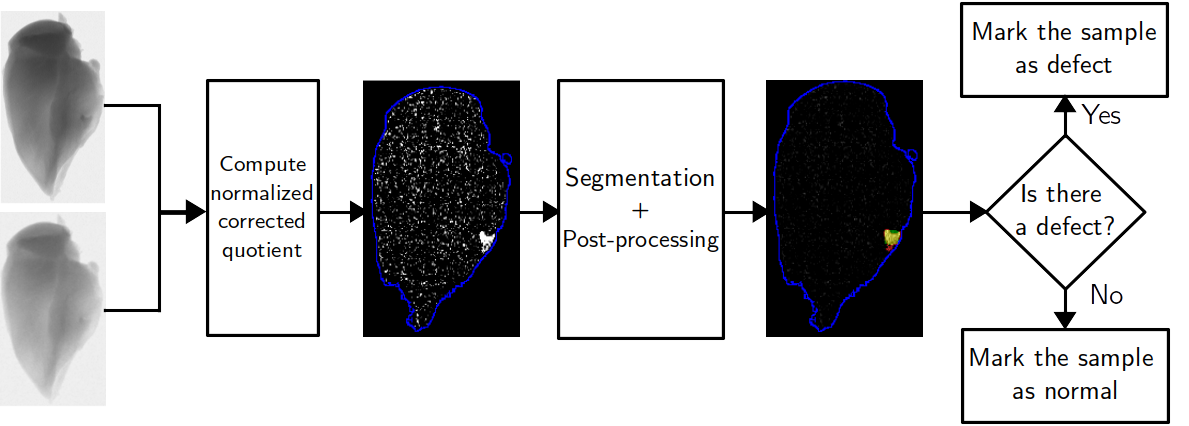}
\caption{Diagram of the foreign object inspection procedure. The input is two projections of the sample acquired with different voltages of the tube. The blue curve on the images approximately shows the sample boundary. The segmentation image uses green color for pixels that were incorrectly classified as a defect, red - for not detected foreign object pixels, and yellow - for detected pixels of the defect.}
\label{flowchart}
\end{figure*}

The product inspection procedure proposed in this paper consists of several stages. The corresponding flowchart is shown on Fig. \ref{flowchart}. Firstly, two X-ray images of the studied sample are obtained using two different voltages of the X-ray tube. These projections should be aligned, and darkfield and flatfield corrected. Then both images are combined into a single quotient using thickness correction pre-processing. The goal of this step is to create an image where pixels of the main object have close to zero values, and a foreign object presence leads to a sufficiently large non-zero intensity. Segmentation is performed on this image to divide it into two phases with different mean intensities. This leads to a set of clusters corresponding to the regions of the foreign object inclusion. Properties of these clusters are used to decide if the sample should be marked as containing a defect.

\subsection{Dual-energy projection pre-processing}

X-ray imaging can be used to create a projection of the studied sample. The value of every pixel in such resulting projections depends on the integral absorption of the object's matter across the corresponding trajectory. The main object and a foreign object absorb radiation differently, and this leads to differences in pixel intensities. However, the shape of the studied sample is not known in advance. Thus, the pixels in the object region do not have constant pixel intensities, as their values depend on the sample thickness. Two images acquired under different voltages provide additional information since material absorption depends on the X-ray photon energy.

A dual-energy projection of a sample with a defect can be segmented as a 2-channel image. The X-ray absorption rate in a pixel depends on both material attenuation properties and the thickness of the object. The absorption rate is given by
\begin{equation}
    M(x) = -\ln \frac{P(x)}{F(x)} = -\ln \frac{\int I_{0}(E) e^{-\kappa(E) L(x)} dE}{\int I_{0}(E) dE},
\label{absorp_eq}
\end{equation}

\noindent where $M(x)$ is absorption rate computed for the detector pixel $x$, $P$ and $F$ are projection and flatfield pixel intensities, $I_0(E)$ is a spectrum of the X-ray tube, $\kappa(E)$ is a material absorption curve, and $L(x)$ is a profile of thickness along the ray. The argument $x$ refers to a detector pixel, and every pixel has a corresponding X-ray beam trajectory from the source to this pixel. The absorption curve $\kappa$ does not depend on $x$ since the material is assumed to be homogeneous. If scattering is not considered, attenuation properties of the material are defined by X-ray absorption, and the attenuation rate can be calculated according to Eq. \ref{absorp_eq}.

If the tube spectrum is monochromatic, then $I_0(E) = I_0 \delta(E - E_0)$, where $\delta(x)$ is a Dirac delta function. Eq. \ref{absorp_eq} can be simplified:
\begin{equation}
    M(x) = -\ln \frac{I_0 e^{-\kappa(E_0)L(x)}}{I_0} = \kappa(E_0) L(x).
\end{equation}

\noindent In this case, the two channels of the dual-energy image are linearly correlated. If a homogeneous material is scanned with two monochromatic beams of energies $E_1$ and $E_2$, the corresponding absorption rates are $M_1(x) = \kappa(E_1) L(x)$ and $M_2(x) = \kappa(E_2) L(x)$. The ratio between $M_1$ and $M_2$ is constant, does not depend on the thickness and is defined by the ratio of attenuation coefficients for two X-ray energies. As a result, two different materials can be easily separated using a dual-energy projection.

In most CT applications, a beam is usually polychromatic since it is produced by conventional X-ray tubes. In this case, the attenuation rate depends on material thickness according to Eq. \ref{absorp_eq}. If the thickness $L(x)$ is small, an effective attenuation coefficient $\kappa_{\textrm{eff}}$ can be computed as a first-order approximation\cite{heismann2003density}:
\begin{equation}
    \kappa_{\textrm{eff}} = \frac{\int I_0(E) \kappa(E) dE}{\int I_0(E) dE}
\end{equation}

However, the attenuation rate does not linearly depend on $L(x)$ in general. Thus, a ratio of attenuation rates is no longer a material characteristic, it depends on the thickness $L(x)$.

\begin{figure*}[]
\centering
\includegraphics[width=0.9\linewidth]{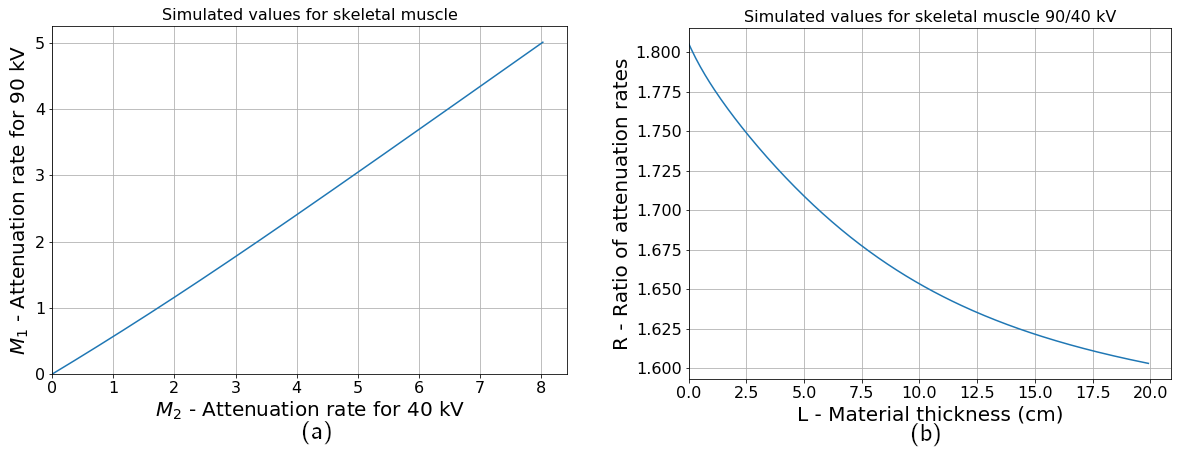}
\caption{Correlation between absorption of skeletal muscle for the X-ray tube voltages of 40 and 90 kV(a). The ratio between the attenuation rate is drawn as a function of thickness. The quotient is not constant due to a polychromatic spectrum (b).}
\label{nist_plots}
\end{figure*}

A nonlinear dependency of attenuation rate on material thickness is visible on the simulated data. For the simulation, the main object is assumed to be a skeletal muscle with an attenuation curve taken from the NIST database (ICRU-44 report\cite{griffiths1989tissue}). The tungsten tube spectrum for the voltages of 40 and 90 kV is computed according to the TASMIP data\cite{boone1997accurate}. Fig. \ref{nist_plots}a shows how attenuation rates for two different voltages of the tube correspond to each other. On this plot, thickness changes from 0.1~mm to 20 cm, and the attenuation rates are calculated according to Eq. \ref{absorp_eq}. The correlation between the two values is almost linear as if both of them depend linearly on thickness. However, the ratio of two attenuation rates changes with thickness, as shown on Fig.\ref{nist_plots}b. This change is not significant, therefore, it is not visible on a correlation plot between two intensities for different voltages. The ratio dependency on thickness can be calculated as
\begin{equation}
    \frac{M_1(x)}{M_2(x)} = \frac{-\ln \int I_{1}(E) e^{-\kappa(E) L(x)} dE + \ln \int I_1(E) dE}{-\ln \int I_{2}(E) e^{-\kappa(E) L(x)} dE + \ln \int I_2(E) dE},
\end{equation}
where $I_1(E)$ and $I_2{E}$ are the tube spectra for two different voltages.

In the real scan, the nonlinear dependency is further complicated by the noise presence. A mass attenuation function is usually unknown for many food industry materials. Therefore, the thickness dependency of the quotient values can not be predicted beforehand and should be extracted from the data. Experimental measurement produces distributions of $M_1(x)$ and $M_2(x)$ for two different tube voltages. The quotient distribution $R(x)$ can be computed as 
\begin{equation}
    R(x) = \frac{M_1(x)}{M_2(x)},
    \label{quotient_formula}
\end{equation}
and the thickness profile $L(x)$ is unknown. As shown on Fig. \ref{nist_plots}a, attenuation rate $M(x)$ is almost proportional to $L(x)$. Thus, in a data-driven approach, the dependency of quotient values on thickness can be studied as a dependency of $R(x)$ on $M_2(x)$. It will be further replaced by a polynomial approximation since a high noise level makes it impossible to recover true dependency $R(M_2)$ from the data without any additional information.

The order of the polynomial chosen for a function approximation depends on the data quality. In the experimental data used in this work, a linear approximation of $R(M_2)$ is not sufficient and leads to significant discrepancies between the acquired data and the fit. High orders of the polynomial are prone to noise, and the fit does not always converge as a result. The quadratic approximation was chosen as a middle ground since it provides a sufficiently good representation of the data and has a low noise sensitivity. This approximation is given by
\begin{equation}
    R(x) = \frac{M_1(x)}{M_2(x)} \approx a M_2^2(x) + b M_2(x) + c,
    \label{ratio_fit}
\end{equation}
\noindent where $M_1(x)$ and $M_2(x)$ are pixel values in the respective channels of the experimentally acquired projection, $a$, $b$, and $c$ are fit coefficients. The polynomial regression is performed for all pixels of the object simultaneously.

When the dependency of $R(x)$ on $M_2(x)$ is extracted from the data in the form of polynomial approximation, the effect of thickness dependency can be reduced. After a polynomial fit, the distribution of $R'(x)$ can be computed as
\begin{equation}
    R'(x) = R(x) - a M_2^2(x) - b M_2(x) - c.
\label{corrected_quotient}
\end{equation}
$R'(x)$ is a corrected quotient distribution. If the sample consists of a homogeneous material, the $R'(x)$ is close to zero regardless of the thickness. However, inclusion of a defect with different absorption properties affects both $R(x)$ and $R'(x)$. $R'(x)$ is easier to use for defect detection since the form variation of the object does not significantly influence this distribution.

\subsection{Pre-processing of the experimental data}

A sample of a chicken fillet containing a fan bone was scanned using a CMOS detector with a CsI(Tl) scintillator (Dexela1512NDT)\cite{Flexray2020}. The X-ray source was a microfocus X-ray tube with voltages of 40 and 90 kV. The piece of fillet was wrapped in a plastic bag and placed on a holder. This experimental setup imitates a top view similar to the typical data from a conveyor belt.

\begin{figure*}[]
\centering
\includegraphics[width=0.75\linewidth]{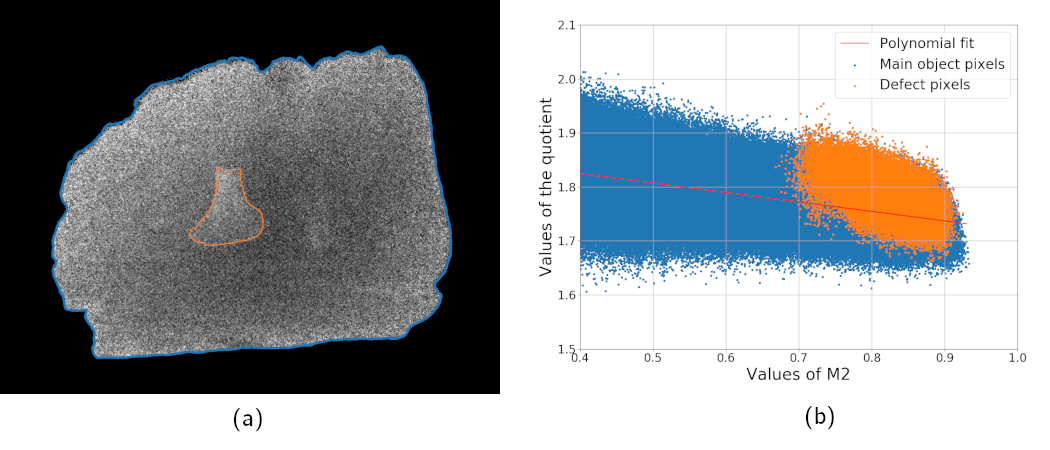}
\caption{Sample scan with low exposure (0.5~s per projection): quotient image computed as a division of two projections (a) and the dependency of quotient values on the single projection intensity (b).}
\label{dual_lowexp}
\end{figure*}

\begin{figure*}[]
\centering
\includegraphics[width=0.75\linewidth]{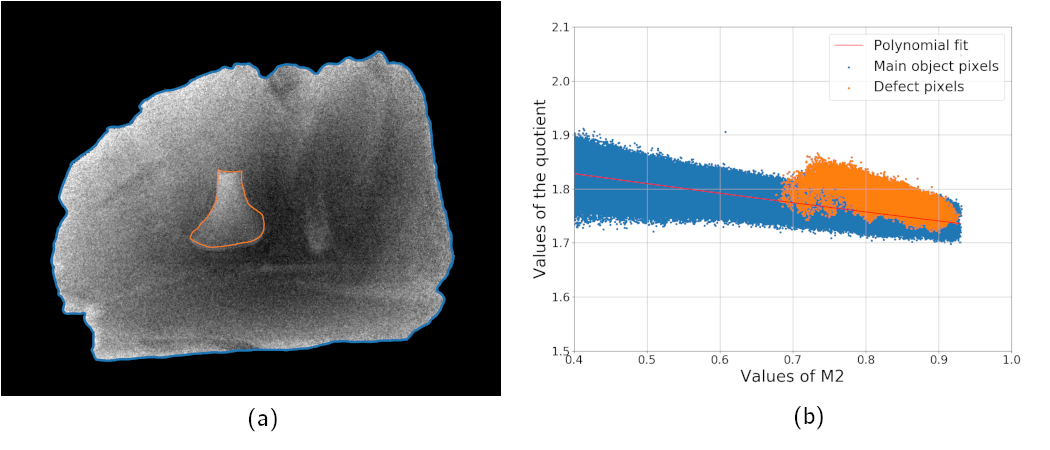}
\caption{Sample scan with high exposure (5~s per projection): quotient image computed as a division of two projections (a) and the dependency of quotient values on the single projection intensity (b).}
\label{dual_highexp}
\end{figure*}

The same sample was measured with different exposure times to illustrate the impact of the detector noise. Fig. \ref{dual_lowexp}a shows a quotient image computed as a division of two projections acquired with the exposure time of 0.5 seconds. Fig. \ref{dual_lowexp}b is a plot of thickness dependency of quotient values based on the experimental data corresponding to Fig. \ref{nist_plots}b for the simulated data. Values of $M_2(x)$ are used instead of $L(x)$ since the thickness profile of the object is unknown. Pixels of the defect are marked with a different color to highlight that the noise variance is bigger than a difference between the sample and defect in spectral properties. Nevertheless, the bone can be located by a human expert based on the quotient image since the defect pixels are located near each other and form a region. If the same product is scanned with a higher exposure time, the level of statistical noise becomes lower, and the foreign object is easier to locate. Figs. \ref{dual_highexp}a and \ref{dual_highexp}b show the quotient image and quotient plot for the measurement with an exposure time of 5~s. The high noise case is more difficult, and it will be the main focus of the next subsections. 

\begin{figure*}[]
\centering
\includegraphics[width=0.9\linewidth]{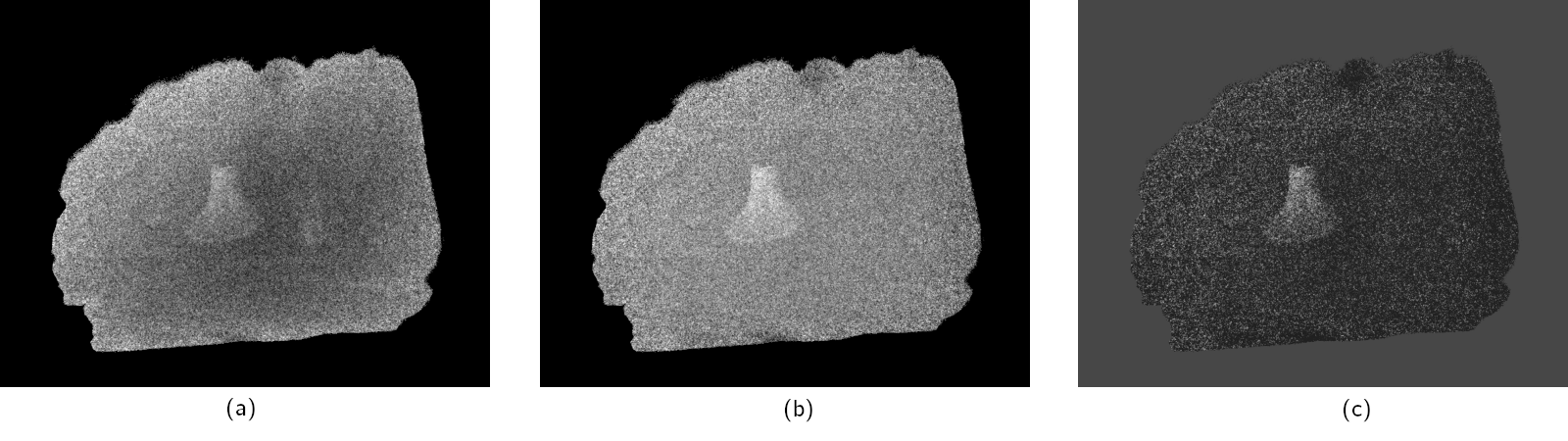}
\caption{Stages of the scan data pre-processing: quotient image computed as a division of 2 channels as in Eq. \ref{quotient_formula} (a), quotient after thickness correction defined by Eq. \ref{corrected_quotient} (b), quotient after correction and normalization computed according to Eq. \ref{normalied_quotient} (c). The sample is a chicken fillet with a fan bone scanned with an exposure time of 0.5~s.}
\label{pre-processing}
\end{figure*}

Even if the foreign object is visible to a human expert, its detection might still be non-trivial for a classical algorithm. Thickness dependency of the ratio values leads to natural gradients in the quotient image. The introduction of the corrected quotient $R'(x)$ helps to reduce this effect and makes the image easier to segment. The thickness correction is shown on Figs. \ref{pre-processing}a and \ref{pre-processing}b that represent a quotient and corrected image. After this procedure, the quotient value is close to zero in most pixels corresponding to the main object. Deviation from zero might be caused by detector noise, systematic errors of the experimental setup, and different defects. The presence of a foreign object changes a pixel value, and the difference depends on its thickness. The main task of the segmentation algorithm applied to Fig. \ref{pre-processing}b is to locate big clusters of nonzero pixels excluding noisy outliers. With a significant noise influence, foreign object location is difficult to perform on a pixel level. Thus, it is important to use spatial information.

The largest noise level in the quotient image is usually found near the edges of the main object. In those regions, a quotient intensity is calculated as a ratio between two small numbers that leads to the high relative error. This effect can be reduced to improve detection accuracy. It can be assumed that the variance of the image values is mainly defined by the statistical noise and depends on the thickness. Therefore, a standard deviation for different thickness values should be computed from the data. The simplest approach to solve this problem is to divide the image into several regions corresponding to different bins of the $M_2$ intensity. If the intensity values start from $m$ and the bin size is $\Delta$, the intensity bins are $[m + i \Delta, m + (i+1) \Delta], i \in \mathbb{N}$. For i-th intensity bin, mean value $\overline{R'_i}$ and standard deviation $\sigma_i$ can be computed for the values of $R'(x), \forall x: M_2(x) \in [m+i\Delta; m+(i+1)\Delta]$. Then the normalized corrected quotient $N(x)$ can be calculated as
\begin{equation}
\begin{aligned}
    N(x) {} &= \frac{R'(x) - \overline{R'_i}}{\sigma_i}, \\
            &\forall x: M_2(x) \in [m+i\Delta; m+(i+1)\Delta].
\end{aligned}
    \label{normalied_quotient}
\end{equation}

Fig. \ref{pre-processing}c shows a normalized quotient based on the corrected image \ref{pre-processing}b.

\subsection{Quotient segmentation}

Thickness correction pre-processing described in the previous section transforms two-channel dual-energy projections into a single normalized quotient. In this work, an active contour model without edges is used to achieve good segmentation quality with a high noise level. The Chan-Vese method is a variational segmentation algorithm inspired by the  Mumford–Shah model. It separates two homogeneous phases in the image by minimizing the energy functional over the phase boundary and their average values
\begin{equation}
\begin{aligned}
    F ={} & \lambda_1 \int_{\Omega_1} (N(x) - c_1)^2 dx + \lambda_2 \int_{\Omega_2} (N(x) - c_2)^2 dx + \\ 
          & + \mu |\partial \Omega_1| + \nu |\Omega_1|,
\end{aligned}
\label{cv_energy}
\end{equation}
where $\Omega_1$ and $\Omega_2$ are regions segmented as an object and backgound, $c_1$ and $c_2$ are average pixel values in these regions, $|\partial \Omega_1|$ is the boundary length of $\Omega_1$, and $|\Omega_1|$ is the area of $\Omega_1$. In the case of a foreign object location problem, the background refers to the main object, and the object refers to the foreign object. The minimization problem is solved by applying the level-set technique: phase boundary is defined as a zero-level of a level-set function. The values of $c_1$ and $c_2$ are recalculated on every step depending on the current phase boundary.

\begin{figure*}[]
\centering
\includegraphics[width=0.9\linewidth]{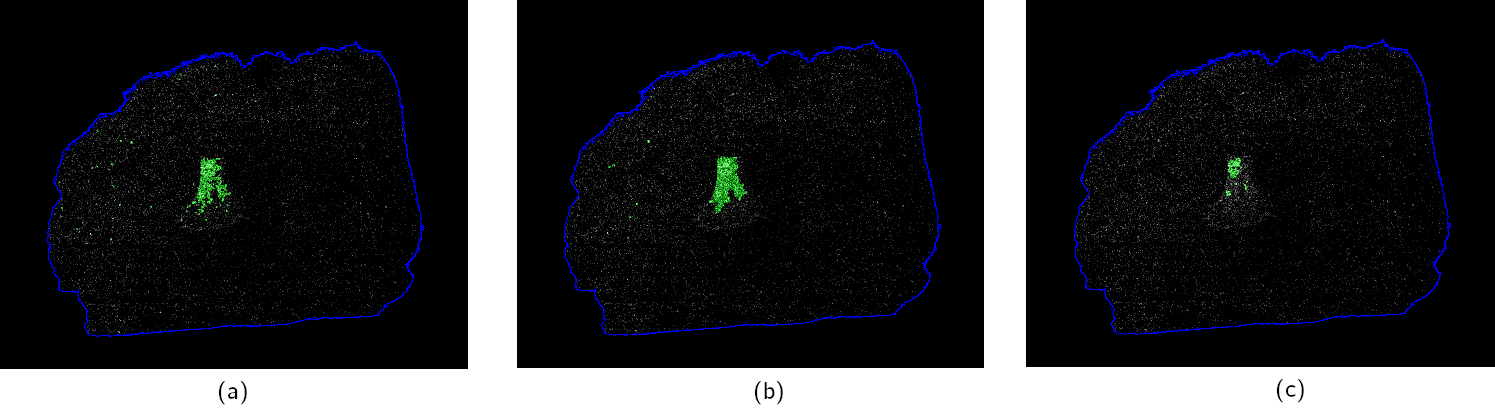}
\caption{Different examples of the segmentation with different values of penalty weights applied to the quotient after pre-processing shown on Fig. \ref{pre-processing}c. The Fig. (a) shows a segmentation with low values of penalty weights $\mu = 1, \nu = 1$ where many noisy outliers are marked as bone fragments. This effect can be reduced by changing the weights as shown on Fig. (b) corresponding to $\mu = 5, \nu = 1$. High values of penalties, such as $\mu = 20, \nu = 5$ on Fig. (c), might lead to a safe segmentation excluding a significant fraction of the bone.}
\label{segmentation_examples}
\end{figure*}

The segmentation outcome is implicitly controlled by the ratios between $\lambda_1$, $\lambda_2$, $\mu$ and $\nu$. The first two terms favor a similarity between pixel value and region average intensity regardless of the spatial properties. The last two terms mitigate the effect of noisy pixels on the segmentation. Low values of $\mu$ and $\nu$ would transform the segmentation into thresholding with minimal removal of outliers (Fig. \ref{segmentation_examples}a). Different values of penalty weights lead to a different boundary detection, noise sensitivity, and overall accuracy of the algorithm. Examples of such effects are shown on Figs. \ref{segmentation_examples}b and \ref{segmentation_examples}c. The biggest strength of the active contour approach is that parameters have an interpretation and can be related to the image properties, such as object intensities and noise distribution.

As a variational algorithm, the Chan-Vese method has a few implicit parameters that influence the iterative process. The initial state of level-set function in the described implementation is defined by a simple thresholding
\begin{equation}
  \phi(x) =
    \begin{cases}
      1 & \text{if}\quad x \geq T_{init}\\
      0 & \text{if}\quad x < T_{init} \\
    \end{cases}.
\end{equation}
where $T_{init}$ corresponds to a significant deviation from zero. Thus, pixels with values higher than the threshold are likely to be part of the defect region. On every iteration of the segmentation algorithm, the level-set is recalculated to better minimize the segmentation energy. If the increment norm is smaller than a certain tolerance value, the algorithm has converged. The iterative process is also stopped if it takes more iterations than a certain maximal number. These parameters mainly influence the speed and convergence of the method and define the final segmentation in case there are multiple local minima.

Accuracy of the segmentation can be evaluated if a ground truth (correct segmentation of the input) is known for every sample. In this paper, F1-score is chosen as an accuracy metrics and is calculated according to the formula
\begin{equation}
    \textrm{F1-score} = \frac{\textrm{TP}}{\textrm{TP} + 0.5 (\textrm{FP} + \textrm{FN})},
    \label{f1_formula}
\end{equation}
\noindent where TP is the number of True Positive pixels (pixels of the defect that were correctly identified), FP - False Positive (pixels of the main object that were falsely classified as a defect), and FN - False Negative (pixels of the defect which were missed). This metric is commonly used in papers about foreign object segmentation. However, it does not evaluate performance on the samples without a foreign object. 

\subsection{Post-processing and foreign object detection}

The main challenge for the active contour segmentation lies in images without defects. A neighborhood of a noisy pixel with a significant deviation from zero can be considered as a part of the defect region, even if there is no foreign object in the sample. This problem can be solved by adjusting penalty weighting coefficients $\mu$ and $\nu$. If a region $\Omega_{noisy}$ with mean intensity $N_{noisy}$ is considered as a part of the main object, the energy will be increased by $(N_{noisy} - c_{main})|\Omega_{noisy}|$ since the mean intensities are different. This energy increment can be avoided if this region is classified as a foreign object. In this case, the energy will be modified by penalty terms $\mu|\partial\Omega_{noisy}| + \nu|\Omega_{noisy}|$. The decision about including or excluding the region $\Omega_{noisy}$ is based on the ratio between these two terms. It is important to note that the Chan-Vese algorithm is an iterative method. Therefore, the result does not only depend on energy terms but on the initial level-set, regularization parameters, convergence speed among other things. Nevertheless, penalty weights should be raised to a certain level to exclude noisy clusters in most cases.

If a noise level is sufficiently high, fine-tuning parameters for all types of inputs is a challenging problem. Significant noise fluctuations in the samples without foreign objects require penalty weights to be high. On the other hand, the accuracy of defect segmentation becomes worse since many pixels on the foreign object boundary are included in the main object. A post-processing procedure is introduced as an additional way to exclude noisy pixels from the foreign object region and make the algorithm more robust. A segmented defect region can be divided into clusters of neighboring pixels. For each cluster, the mean intensity and size can be computed. Segmentation quality can be enhanced if a certain threshold on the cluster size is set, and small clusters are ignored. The main reason for employing this strategy is the existence of noisy pixels with an intensity that is several times higher than the average defect intensity. If penalty weights are adjusted to exclude such pixels, small foreign objects might be excluded as well.

In the proposed methodology, post-processing is the removal of segmented clusters with a size lower than a certain number of pixels. If there are no clusters after pre-processing, the sample is marked as normal. Otherwise, it is considered that the sample contains a defect. For every sample in the experimental dataset, it is known whether it contains a foreign object or not. Thus, it is possible to compute a confusion matrix and F1-score for the inspection procedure. In this case, accuracy is measured on a sample level, unlike the segmentation precision. These metrics are more important for the algorithm performance evaluation since they include all possible cases. If the segmentation is fine-tuned to achieve the best segmentation accuracy, it might become too sensitive to noise. Therefore, due to noise fluctuations, it will classify normal samples as containing a defect.

\section{Results}

\subsection{Dataset description}

The thickness correction procedure was tested on the scans of chicken filets on a conveyor belt. The images were acquired with a line detector since these are commonly used in industrial setups. The majority of filet samples contained a bone that should be detected as a foreign object. Every sample was scanned four times with different positions on a belt. There are 100 images with a fan bone, 100 images with a large rib bone, and 96 images with a small rib bone. In 192 scans a chicken fillet does not contain a foreign object. These types of bone differ by average size, form, and typical position in a fillet. The dataset was semi-manually segmented to create an approximate ground truth for accuracy estimation.

The biggest challenge of the dataset is the small difference in attenuation curves between the chicken fillet and bone. Both are organic materials and contain almost the same chemical elements in different ratios. As a result, the dependency of attenuation coefficients on energy is similar, and therefore, spectral imaging efficiency is limited. The difference in spectral properties is visible but of the same order of magnitude as a noise level.

\begin{table}[]
\caption{Comparison of mean pixel intensity and foreign object sizes for different types of bone}
\label{stat_table}
\begin{tabular}{|l|l|l|}
\hline
Bone class     & Pixel value & Bone size \\ \hline
Fan bone       & 3.7 $\pm$ 2.6   & 370 $\pm$ 120 \\ \hline
Large rib bone & 3.1 $\pm$ 2.5   & 290 $\pm$ 70  \\ \hline
Small rib bone & 3.0 $\pm$ 2.6   & 160 $\pm$ 35  \\ \hline
\end{tabular}
\end{table}

The ground truth for the dataset was used to calculate the average properties of different bone types. Table \ref{stat_table} shows the comparison between pixel values after thickness correction and normalization and bone sizes for different classes. 

\subsection{Thickness correction}

\begin{figure*}[]
\centering
\includegraphics[width=0.8\linewidth]{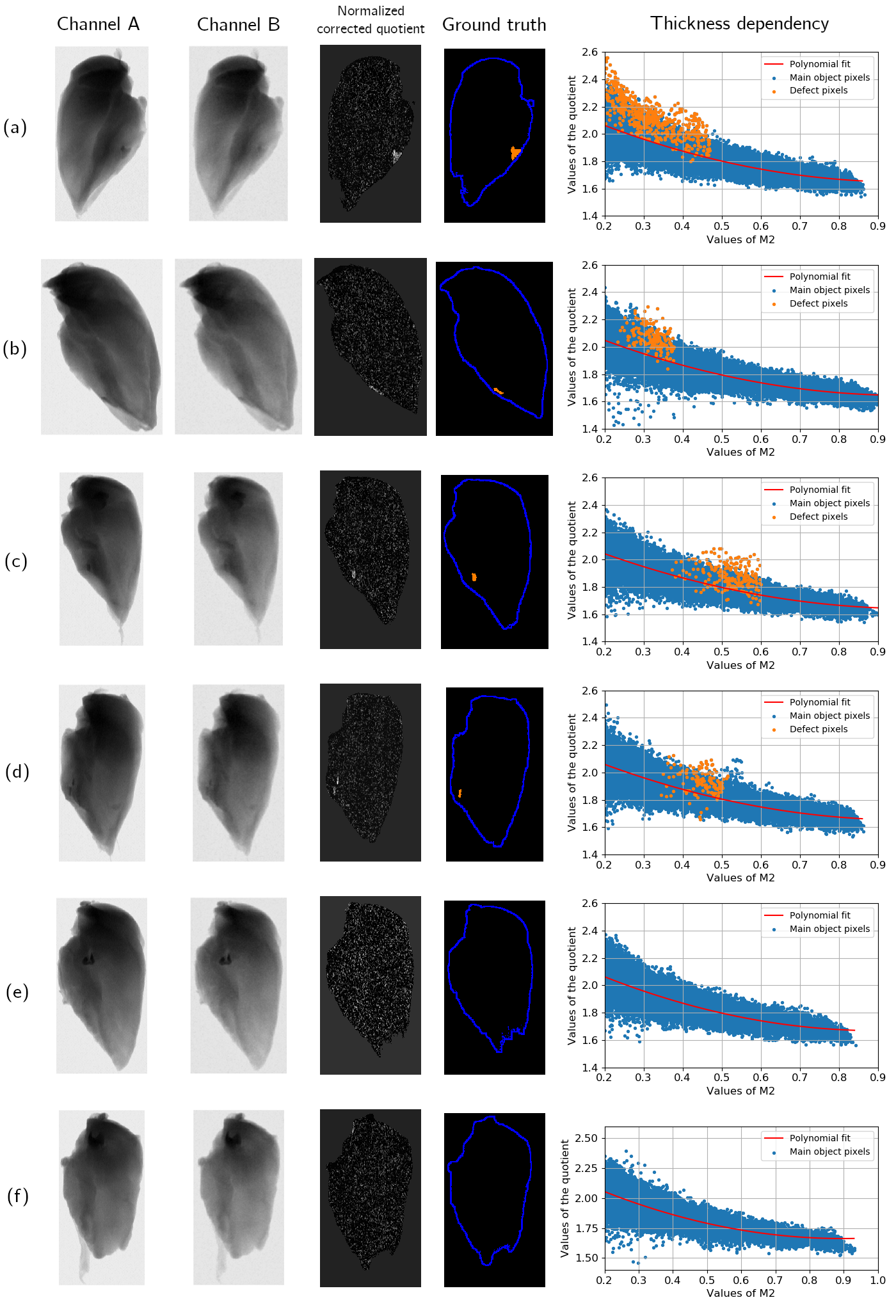}
\caption{Different samples from the experimental dataset. For every object, two projections acquired with different voltages, the normalized corrected quotient, the segmented image, and the thickness dependency plot are shown. Sample (a) contains a fan bone, (b) - large rib bone, cases (c) and (d) show small rib bones. Sample (e) and (f) do not contain defects. Boundaries of the samples are approximately drawn as blue curves but they are not used during the inspection procedure. The defect location is marked by orange color and corresponds to the ground truth images from the dataset. Ground truth in sample (d) is partially wrong and does not include the second part of the shattered bone.}
\label{dataset_examples}
\end{figure*}

The described pre-processing was applied to the dataset to compute normalized corrected quotient images. A thickness dependency for every sample was interpolated by a polynomial of the second degree. If an absorption rate was lower than a certain threshold equal to 0.2, a pixel was ignored both in interpolation and division. The size of the intensity bin $\Delta$ was set to 0.1. Fig. \ref{dataset_examples} shows projections of different samples from the experimental dataset, the corresponding normalized corrected quotient for every case, the ground truth, and the thickness dependency plot.

The dataset contains a variety of examples with different types and sizes of bone fragments and normal samples without defects. Cases (a) - (d) show the effect of thickness correction on samples with a foreign object. The defect might be visible on a single projection, even if it has a small size, such as a shattered rib bone in sample (d). However, the contrast on a single projection significantly depends on the exact location of the foreign object. If the main object form causes intensity gradients near the defect, it might be missed without additional information. The main benefit of thickness correction lies in removing intensity changes corresponding to the main object and highlighting the defect location. The thickness dependency plot shows that quotient values corresponding to the foreign object often have a similar deviation from zero as noisy outliers. This corresponds to the low exposure scanning procedure discussed previously.

Example (e) illustrates the main advantage of using thickness correction for DEXA data. Both projections contain a region with a well-visible border that is not a foreign object. However, the correlation between images does not correspond to a material with attenuation properties different from the main object. In the corrected quotient, this region is the same as other parts of the sample. The thickness dependency plot does not contain any significant outliers as well. Such intensity changes might appear on samples with and without foreign objects. Therefore, it is crucial to remove them in order to prevent a high false positive rate.

Some projections in the dataset contain systematic experimental effects that can look like a foreign object after thickness correction. In sample (f), a set of pixels near the border has a high quotient value after correction. However, no bone is present in the object in this case. Automatic data acquisition might lead to misalignment artifacts, small movements of the object between scans and change of shape. These artifacts have quotient values similar to the real foreign objects and might be recognized as such.

\subsection{Segmentation accuracy}

Normalized quotient images were used as an input for the Chan-Vese segmentation algorithm. The method implementation was based on the C++ code by Pascal Getreuer \cite{getreuer2012chan}. Furthermore, a Python wrapper was written and used as an interface. In the results the default algorithm parameters are fit weights $\lambda_1 = \lambda_2 = 1$, time step $dt = 1$, convergence tolerance $tol = 10^{-4}$, maximal number of iterations $N_{max} = 200$, Heavyside regularization $\epsilon = 1$, curvature regularization $\eta = 10^{-8}$, initial level-set threshold $T_{init} = 5$. Accuracy of the algorithm is studied for different values of $\mu$ and $\nu$ since they significantly influence the outcome and have a geometrical interpretation. The post-processing pixel count threshold is set to 30 pixels. This means that defect clusters containing fewer than 30 neighbouring pixels were removed from the final segmentation.

\begin{figure*}[]
\centering
\includegraphics[width=0.8\linewidth]{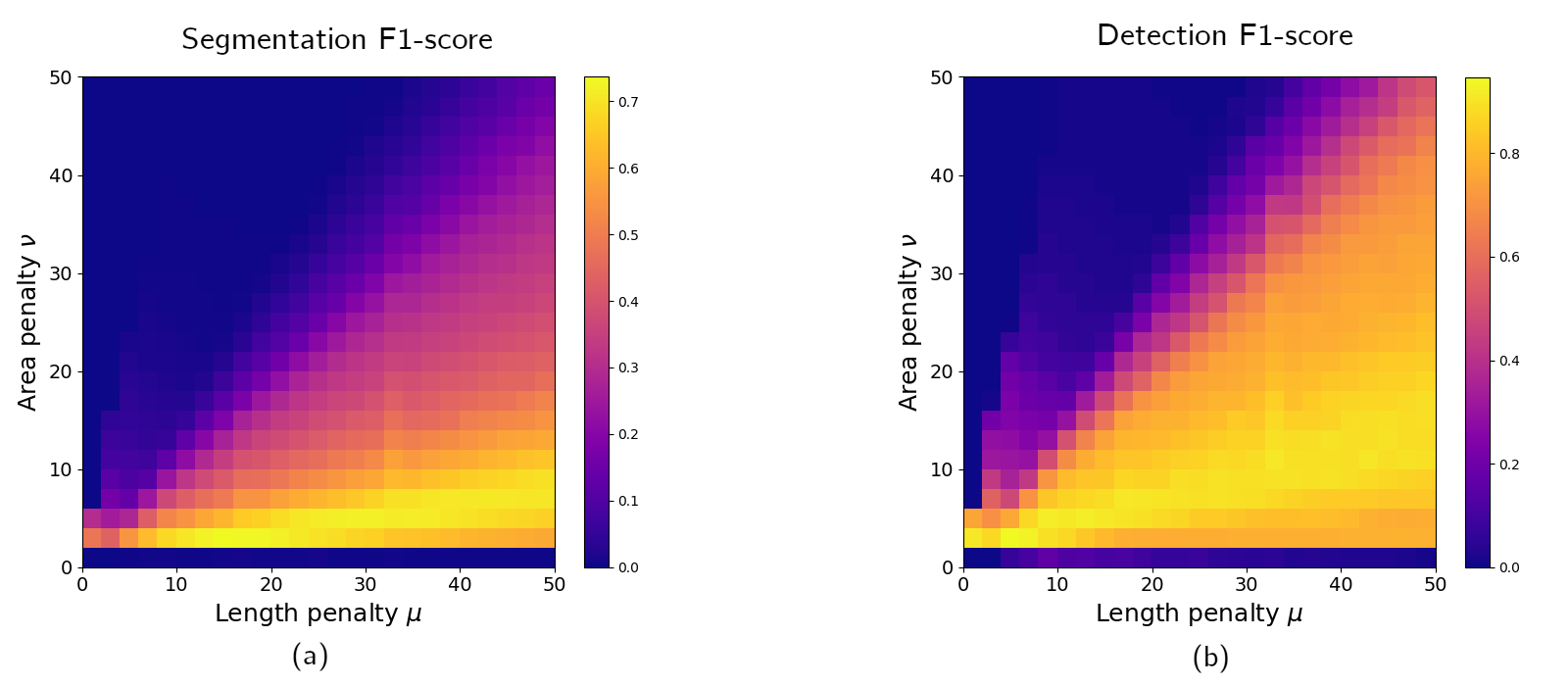}
\caption{Dependency of F1-score on length penalty $\mu$ and area penalty $\nu$ for different tasks: image segmentation (a) and foreign object detection (b). For segmentation, the F1-score is computed using a ground truth segmentation known for every sample and averaged over all images from the dataset containing a defect. For detection, the metric is calculated on a sample level for the entire dataset consisting of the objects with and without a bone.}
\label{f1scores}
\end{figure*}

In this subsection, the segmentation accuracy is estimated on a pixel level for the samples containing a foreign object. For every object, the resulting segmentation is compared with ground truth to count the number of true positive, false positive, and false negative pixels. F1-score is calculated according to Eq. \ref{f1_formula}. The values of the F1-score are shown on Fig.~\ref{f1scores}a for different combinations of penalty weights. The value of the metric is averaged over all images with foreign objects. The best segmentation accuracy is achieved with $\mu=14$ and $\nu=2$.

\begin{figure*}[!t]
\centering
\includegraphics[width=0.9\linewidth]{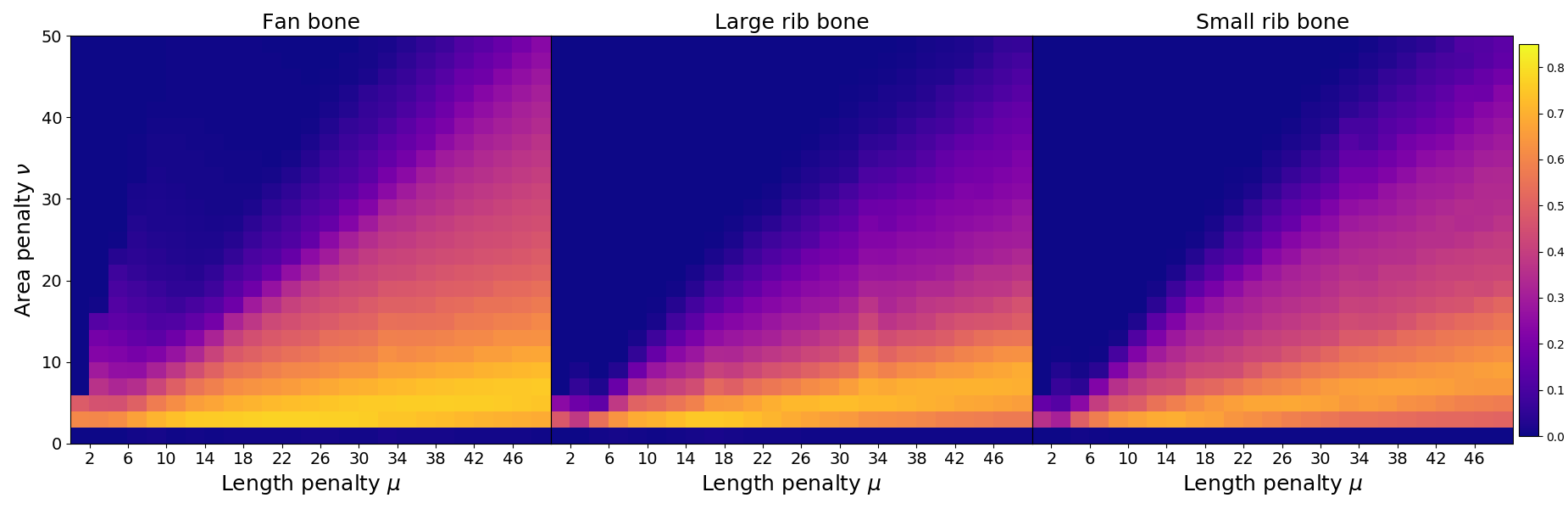}
\caption{Defect segmentation F1-score for every class of defect in the dataset: fan bone, large rib bone, small rib bone. The F1 value is averaged over all samples with a corresponding defect type.}
\label{class_comparison}
\end{figure*}

As explained in Section~4, the dataset contains different types of bone as a defect. Values in Fig.~\ref{f1scores}a are averaged over all defect types. Thus, it does not show how defect class affects segmentation accuracy. The corresponding figures for every type of bone are shown on Fig.~\ref{class_comparison}. The dependencies of the F1-score on penalty weights are similar for all defects. Every bone class has a combination of Chan-Vese parameters that achieves the best segmentation accuracy for that defect type, and these parameters might be different from each other. However, the best instances for a single defect class also perform well for the whole dataset as shown in Table~\ref{comparison_table}. Thus, all types of defects can be segmented with the same algorithm settings. Different ratios of bone types in the dataset would affect the algorithm performance, but not significantly.

\begin{table}[]
\caption{Comparison of the best Chan-Vese parameters for different classes of defects. F1-score is separately calculated for all images with the same bone class and for all samples with a defect.}
\label{comparison_table}
\begin{tabular}{|c|c|c|c|c|}
\hline
Defect class   & $\mu$ & $\nu$ & Single class F1 & F1 for all defect classes \\ \hline
Fan bone       & 20 & 2  & 77$\%$                           & 72$\%$                            \\ \hline
Large rib bone & 14 & 2  & 75$\%$                           & 74$\%$                            \\ \hline
Small rib bone & 14 & 2  & 70$\%$                           & 74$\%$                            \\ \hline
\end{tabular}
\end{table}

\subsection{Detection rate}

As highlighted previously, the detection procedure should be evaluated on the samples without a foreign object. The decision-making based on the segmentation and post-processing is tested on the whole dataset: 296 images with different types of bone and 192 images without a defect. The test results contain the number of images with a bone where a presence of defect was detected and the number of boneless images that were correctly identified as empty. These values correspond to the true positive and true negative cases. The algorithm accuracy is evaluated using the F1-score.

Fig. \ref{f1scores}b shows the dependency of F1-score on Chan-Vese energy equation penalties $\mu$ and $\nu$. High accuracy(more than 90\%) can be obtained with multiple combinations of parameters, the best value of F1-score is achieved with $\mu=4$ and $\nu=2$. The confusion matrix for this instance of the inspection procedure is shown in Table \ref{confusion_matrix}. The algorithm correctly marks 97\% of the normal samples as not containing a foreign object. Chicken fillets with a bone are successfully identified in 92\% of cases. On average, segmentations with the best detection rate converge after 150 ms.

{
\begin{table}
\caption{Confusion matrix for the inspection procedure with Chan-Vese parameters $\mu=4$ and $\nu=2$}
\label{confusion_matrix}
\begin{tabular}{@{}cc cc@{}}
\multicolumn{1}{c}{} &\multicolumn{1}{c}{} &\multicolumn{2}{c}{Predicted} \\ 
\multicolumn{1}{c}{} & 
\multicolumn{1}{c}{} & 
\multicolumn{1}{c}{Defected} & 
\multicolumn{1}{c}{Normal} \\ 
\cline{2-4}
\multirow[c]{2}{*}{\rotatebox[origin=tr]{90}{Actual}}
& Defected  & 271 & 25   \\[1.5ex]
& Normal  & 5   & 187 \\ 
\cline{2-4}
\end{tabular}
\end{table}
}

\section{Discussion}

\subsection{Thickness correction}

The thickness correction pre-processing is performed on the experimental data and does not rely on prior knowledge about the samples. The dependency of the quotient value on the thickness of the main object material is estimated as an average function for pixels of the projection. The theoretical foundation of the pre-processing implies that this function can be predicted with sufficient knowledge of the inspection system. As a result, it can be used to construct more precise measurement systems and achieve better contrast between the main and foreign objects. However, in the scope of this article, a heuristic approach was chosen to show the applicability of this method to a wide range of data. 

One of the major downsides of the quotient images after thickness correction is the resulting significant level of noise. If noise in single-energy projections follows a Gaussian distribution, the quotient noise has a ratio distribution. Thus, significant outliers from the mean value are more likely than with the Gaussian. The exact properties of the resulting distribution depend on the mean value and variance for both original distributions. In practice, this means that high noise appears frequently, especially in the boundary regions of the image.

Several approaches were considered to mitigate a high noise level. Firstly, normalization is applied to the corrected quotient. On the boundary, big deviations from zero level are divided by the significant variance value. A downside of this procedure is that variances are computed under the assumption that the distribution is Gaussian. Secondly, not all boundary regions are used for the quotient computations. If a pixel value is less than a certain threshold, a pixel is considered to be a part of the background and ignored. This boundary cut might remove some parts of the bone from the quotient if it is located near the main object border. Nevertheless, a smaller number of noise outliers leads to a better segmentation in general.

In this work, the normalized corrected quotient images are used as an input for a segmentation algorithm. However, it does not mean that other derivatives of two projection channels should not be considered. In the data-driven approach, the quotient is used as a simple combination of two channels that has an additional meaning for a monochromatic beam. With more knowledge about the inspection system and product materials, a better combination of the two channels might be constructed. The main focus of the pre-processing procedure is to remove thickness dependency, additional steps can be considered to improve defect contrast. Normalization of the quotient can be viewed as an implicit introduction of the Gaussian noise to the active contour model.

\subsection{Active contour segmentation}

The Chan-Vese method operates well even with noisy data, and a high noise level is inevitable for the conveyor belt product inspection. The energy that is minimized over two phases in the image is a formal way of defining a connected cluster of points in the presence of high noise. Thus, the segmentation algorithm determines bone borders consistently based on the objective criteria. At the same time, ground truth made by a human operator might be more subjective. When a discrepancy between the segmentation and ground truth happens, it can be caused by many factors. On one hand, the Chan-Vese method might not converge or reach a local minimum, and thickness correction might produce a very noisy input. On the other hand, the ground truth in a single sample can be inconsistent with other data.

The active contour models contain several parameters that do not have a physical meaning. The maximal number of iterations, convergence tolerance, and time step influence the speed of the method and resulting segmentation. The optimal choices of these parameters balance computational speed and algorithm accuracy. For the detection procedure, it is crucial to get the result as fast as possible. Therefore, a large time step and low tolerance can be considered.

Fig. \ref{f1scores} shows that good accuracy can be achieved with a range of algorithm parameters. The best penalty weight pair does not lead to a significantly better F1-score than its neighborhood in the parameter space. Thus, a search for the optimal inspection settings converges quickly. In these plots, the grid step for $\mu$ and $\nu$ was set to 2. A smaller step was not chosen to prevent overfitting to the experimental data. The form of the F1-score dependency on $\mu$ and $\nu$ implies that a similar result can be achieved with a broader dataset.

\subsection{Defect detection}

The active contour model in the described methodology is defined for two homogeneous phases according to Eq. \ref{cv_energy}. This means that the method is well-suited for the cases when a single defect is present in the main object. Two foreign objects of significantly different classes (i.e. bone and a plastic piece) might be segmented incorrectly since they should be marked as a defect phase, but their average intensities vary significantly. This problem can be solved with a change of the active contour energy and the introduction of several level-set functions. However, in practice, it is unlikely that multiple defects appear on the sample since a single foreign object is expected to appear rarely. At the same time, the introduction of more defect types in the energy equation might decrease detection accuracy even more since the majority of samples contains no foreign objects. In the experimental dataset, some images contain a shattered bone, and both pieces are segmented properly since they correspond to the same defect class.

The Chan-Vese algorithm was chosen as a segmentation method in this paper because it performed better than other well-known techniques, such as thresholding and watershed algorithms. However, supervised machine learning methods\cite{ruger2019automated} can be considered as an alternative to classical algorithms. With a relatively small dataset, such as one used in this study, it is possible to train a neural network that is achieving comparable and sometimes better accuracy. Nevertheless, unsupervised algorithms provide explainable results and can be used to improve the experimental setup. With the thickness correction pre-processing, every pixel value can be computed using physical properties of materials, tube parameters, and detector model. The penalty weights can be interpreted as a balance between the defect signal and noise level of the image. This information can be used to improve the scanning protocol, evaluate the cost efficiency of different detectors for a certain task, and estimate the size limits of the detectable foreign objects. Machine learning lacks this level of result explainability and should be applied to an already optimal experimental setup.

The described methodology is not limited to the food industry. The main novelty of this inspection procedure is the thickness correction procedure. The effect of thickness on quotient values is relevant for any dual-energy single projection measurement. It is not necessary if the defect has significantly different attenuation properties (e.g. detection of metal pieces in the luggage). Nevertheless, the thickness is important to account for if the data contain high noise level and low-contrast foreign objects.

The detection task with optimal parameters achieves 95\% accuracy on the experimental dataset. 5 samples out of 192 were misclassified as containing a bone. Some of them can be attributed to systematic experimental errors, such as misalignment or sample deformation. In other cases, a noisy cluster is not segmented as a defect if the convergence tolerance and the maximal number of iterations are changed. In 25 samples out of 296, a bone is not detected. It is theoretically possible to achieve 100\% accuracy for defect detection on the limited dataset. The main limiting factor is detector noise that requires strict length and area penalties for the segmentation. Furthermore, different materials present in the factory environment, such as blood droplets or poultry fat, can be recognized as a foreign object and lead to the false positive case.

It is important to note that the best Chan-Vese parameters for defect detection are different from those that provide the best segmentation accuracy. For a binary outcome, it is not important how precisely the bone is located on the image. The segmentation method often marks only a central part of the bone and ignores its boundary. At the same time, the best detection parameters lead to better performance in difficult cases: the presence of small bones that are indistinguishable from noise and significant noise fluctuations that look similar to small bones. Furthermore, the execution time is lower for the detection procedure which is important in the industrial environment.

\section{Conclusion}
Thickness correction pre-processing proposed in this work enables accurate detection of foreign objects in the dual-energy projections of the conveyor belt samples. The described methodology does not rely on a good contrast between a defect and the main object on a single projection. Instead, it utilizes the difference in attenuation properties that can be detected with a dual-energy acquisition. The active contour segmentation algorithm allows analyzing data with a significant noise level if a proper energy model is chosen. The performance of the inspection is evaluated based on the ability to distinguish samples with and without a foreign object. It was shown that 97\% of samples without a defect can be correctly identified while maintaining a 95\% accuracy of the foreign object detection on the experimental dataset. The proposed approach does not require prior knowledge about the samples, and necessary material properties are extracted directly from the projections. The methodology is tested on bone detection in chicken filets. However, the thickness correction procedure does not rely on any specific properties of this problem and can be extended for other foreign object detection tasks. Different reasons for segmentation imperfection and possible ways to improve the current implementation are discussed.

\section*{Acknowledgment}
The authors acknowledge financial support from the Netherlands Organisation for Scientific Research (NWO), project number 639.073.506. The authors would like to thank Meyn Food Processing Technology B.V. for providing the experimental dataset and Sophia Bethany Coban for assistance with laboratory data acquisition.


%


\ifCLASSOPTIONcaptionsoff
  \newpage
\fi



%
\bibliography{main.bib}
\bibliographystyle{IEEEtran}

%








\end{document}